%% file: uq-and-instabilities.tex
\documentclass[runningheads]{llncs}
\usepackage{graphicx}
\usepackage[english]{babel}
\usepackage{microtype}
\usepackage{amsmath}
\usepackage{commands/nkj}
\usepackage{multirow}
\usepackage{booktabs}
\usepackage{enumitem}
\usepackage{hyperref}
\usepackage{cleveref}
\usepackage{todonotes}
\usepackage{tikz}
\usetikzlibrary{spy}
\usepackage{float}
\usepackage[utf8]{inputenc}

\input{commands/abbrv}

\newcommand{\DropLong}{Monte Carlo Dropout}
\newcommand{\ProbLong}{Mean \& Variance Estimation}
\newcommand{\IntervalLong}{Interval Neural Network}
\newcommand{\DropShort}{\textsc{MCDrop}}
\newcommand{\ProbShort}{\textsc{ProbOut}}
\newcommand{\IntervalShort}{\textsc{INN}}

\newcommand{\CTLong}{Computed Tomography Reconstruction}

\newcommand{\CTShort}{\textsc{CT}}

\newcommand{\ChestCT}{Limited Angle CT}
\newcommand{\Denoise}{Image Denoising}

\newcommand{\AdvLong}{Adversarial Artifact Detection}
\newcommand{\ArtLong}{Atypical Artifact Detection}
\newcommand{\AdvShort}{AdvDetect}
\newcommand{\ArtShort}{ArtDetect}

\newcommand{\repeatthanks}{\textsuperscript{\thefootnote}}

\begin{document}
\title{Interval Neural Networks as Instability Detectors for Image Reconstructions}
\titlerunning{INNs as Instability Detectors}
%

\author{Jan Macdonald\thanks{Equal contribution in alphabetical order}\inst{1} \and
Maximilian März\repeatthanks\inst{1} \and
Luis Oala\repeatthanks\inst{2}\and
Wojciech Samek\inst{2}}
\authorrunning{J. Macdonald et al.}

\institute{
Dept. of Mathematics, Technical University of Berlin, 10623 Berlin, Germany\\
\email{\{maerz,macdonald\}@math.tu-berlin.de} \and
Machine Learning Group, Fraunhofer HHI,
10587 Berlin, Germany\\
\email{\{luis.oala,wojciech.samek\}@hhi.fraunhofer.de}}

\maketitle
%
%
\begin{abstract}

This work investigates the detection of instabilities that may occur when utilizing deep learning models for image reconstruction tasks. Although neural networks often empirically outperform traditional reconstruction methods, their usage for sensitive medical applications remains controversial. Indeed, in a recent series of works, it has been demonstrated that deep learning approaches are susceptible to various types of instabilities, caused for instance by adversarial noise or out-of-distribution features. It is argued that this phenomenon can be observed regardless of the underlying architecture and that there is no easy remedy. Based on this insight, the present work demonstrates on two use cases how uncertainty quantification methods can be employed as instability detectors. In particular, it is shown that the recently proposed \emph{\IntervalLong s} are highly effective in revealing instabilities of reconstructions. Such an ability is crucial to ensure a safe use of deep learning-based methods for medical image reconstruction.    

\end{abstract}
%
%
\section{Introduction}
\label{sec:introduction}
Deep learning has shown the potential to outperform traditional schemes for solving various signal recovery problems in medical imaging applications \cite{kang2017,jin_deep_2017,Hammernik2018,adler2018}.
Typically, such tasks are modelled as finite-dimensional linear inverse problems,
\begin{equation}\label{eq:inv_prob}
 \textstyle\bfy = \bfA \bfx + \bfeta,
\end{equation}
where $\bfx\in\R^n$ is the unknown signal of interest, $\bfA\in\R^{m\times n}$ denotes the forward operator representing a physical measurement process, and $\bfeta\in\R^m$ is modelling noise in the measurements. Important examples include choosing $\bfA$ as the identity (denoising), a subsampled Fourier matrix (magnetic resonance imaging), or a discrete Radon transform (computed tomography).  
Solving the inverse problem \eqref{eq:inv_prob} amounts to computing an approximate reconstruction of $\bfx$ from its observed measurements $\bfy$. The difficulty of this task is mainly determined by the strength of the noise and the degree of ill-posedness of \eqref{eq:inv_prob}, which is typically governed by the amount of undersampling in the measurement domain; cf.~\cite{hansen2010invprobs,Foucart2013}. 

In many cases, sparse regularization provides state-of-the-art solvers for~\eqref{eq:inv_prob}, which are additionally backed up by theoretical guarantees, e.g.~by compressed sensing~\cite{Foucart2013}. Recently, it has been demonstrated that data-based deep learning methods are able to outperform their traditional counterparts in terms of empirical reconstruction quality and speed, however, the field is still in an early stage. Focusing primarily on recovery performance, aspects such as the reliability of reconstructions have not yet been extensively explored; see \cite{adler_deep_2018,ardizzone2018} for exceptions. 

In image classification, the susceptibility of deep neural networks to adversarial exploitation is well documented \cite{42503,moosavi,10.1145/3128572.3140444}. Recent works have reported similar instabilities for image reconstruction tasks \cite{huang2018,vegard2019,gottschling2020}, which can be caused by visually imperceptible adversarial noise or features that have not been seen during training. Although there have been first attempts to alleviate these shortcomings~\cite{Raj2020,Bubba_2019}, \cite{gottschling2020} argues that such instabilities are in fact an unavoidable price for improvements in performance over classical methods. Hence, this work is motivated by the following premise: \textit{if instabilities occur, we want to be able to detect them}. To that end, we demonstrate the potential of uncertainty quantification (UQ) as an instability detector. Out of the three compared UQ methods, the recently proposed \IntervalLong\ framework of \cite{INN} is shown to be particularly well suited for this task. 

\subsubsection{Overview and Contributions}
We consider a straight-forward approach to solving \eqref{eq:inv_prob}, which is based on post-processing a standard model-based inversion by a neural network~\cite{zhang2016,kang2017,jin_deep_2017}. Thus, the reconstruction is given by
\begin{equation}
\label{eq:inv_method}
\textstyle\bfx_{\text{rec}} = \bfPhi(\bfA^\dagger\bfy),
\end{equation}
where $\bfPhi\colon\R^n\to\R^n$ denotes the prediction network (trained to minimize the loss $\| \bfx-\bfPhi(\bfA^\dagger\bfy)\|_2^2$) and $\bfA^\dagger$ symbolizes the non-learned model-based inversion.\footnote{There is a variety of other possibilities to utilize neural networks to solve \eqref{eq:inv_prob}; see~\cite{arridge_maass_oektem_schoenlieb_2019} for a comprehensive overview. However, \cite{vegard2019} suggests, that the issue of instabilities occurs independently of the considered reconstruction scheme. Thus, we restrict our study to the simple "image-to-image" post-processing setting described above.} 
Based on this reconstruction method, we then focus on two use cases. First, the standard imaging task of removing white Gaussian noise (i.e., $\bfA$ is the identity), which can be seen as a well-conditioned inverse problem, is examined.\footnote{Note that there is an intimate connection between denoising and solving general ill-posed inverse problems, that can for instance be exploited by ``plug-and-play'' schemes~\cite{venkat2013}; see also~\cite{oymak}. Thus, this application is chosen as a prototypical example for image-to-image regression by neural networks with a broad scope of implications.}
Second, we consider the severely ill-posed problem of limited angle computed tomography ($\bfA$ is a subsampled Radon transform), which has applications in dental tomography, breast tomosynthesis or electron tomography. While $\bfPhi$ is only used for a plain removal of Gaussian noise in the first case, the latter application requires a removal of structured artifacts as well as an ``inpainting" of missing edge information. On each of the two use cases, we investigate the capacity of three UQ schemes (see Section~\ref{sec:methods}) to localize possible instabilities in the output of the prediction network $\bfPhi$. As possible causes for such instabilities we consider: (i) adversarial noise on the input and (ii) imposed structural characteristics that have not been seen during training, i.e., out-of-distribution (OoD) features (see Section~\ref{sec:main}). We believe that detecting OoD-instabilities is of particular importance in the context of medical imaging, since pathological changes are typically rare events in the training data.   

In summary, the contributions of this work are as follows:
\begin{enumerate}
    \item[\textbf{a)}] We show that UQ can be utilized to detect the lack of robustness of deep learning-based image reconstruction methods.
    \item[\textbf{b)}] Three UQ schemes for artificial neural networks are compared with respect to their capacity of revealing reconstruction instabilities described by \cite{gottschling2020,vegard2019,huang2018}.
    \item[\textbf{c)}] We demonstrate that one UQ approach in particular, the so called \IntervalLong, performs best as an instability detector.
\end{enumerate}
\subsubsection{Related Work}
\label{sec:relatedwork}
In addition to the work cited above there exist strands of research in deep learning occupied with the detection of adversarial and OoD inputs. Maximum Mean Discrepancy, Kernel Density Estimation and other tools, see \cite{10.1145/3128572.3140444} for an overview, have been successfully employed for adversarial input detection. Popular methods for OoD input detection include Minimum Covariance Determinant \cite{doi:10.1080/01621459.1984.10477105}, Support Vector Data Description \cite{10.1023/B:MACH.0000008084.60811.49}, as well as methods geared particularly towards the deep model setting such as ODIN \cite{LiangLS17},  Outlier Exposure \cite{hendrycks2019oe}, or detection in latent space \cite{doi:10.1021/acscentsci.7b00572}.

The detection of adversarial and OoD inputs in these works is typically done in the classification setting. We emphasize that image-to-image regression by $\bfPhi$ is a fundamentally different task: While classification is inherently discontinuous, $\bfPhi$ addresses a problem  that allows for stable reconstruction methods in many cases, e.g.~by sparse regularization. Furthermore, we are not interested in a crude, outright rejection of data points in the \textit{input space} but rather seek to obtain fine-grained information about erroneous artifacts in the \textit{output space}. More closely related to our goal is the work of \cite{kendall_what_2017,gast_lightweight_2018} where uncertainty quantification was considered for segmentation and depth-estimation tasks. Hence, we include their approaches as detection methods which are described next.
\section{Detection Methods}
\label{sec:methods}
We consider three methods for uncertainty quantification of neural network predictions and compare their capacity to detect reconstruction instabilities caused by adversarial noise and OoD features.
\subsubsection{\IntervalLong}
By using interval arithmetic a baseline network $\bfPhi\colon\R^n\to\R^n$ can be extended to an \IntervalLong\ (\IntervalShort)
\begin{equation}\label{eq:inn}
\textstyle\bfPhi_\text{\IntervalShort}\colon\R^n\to\R^n\times\R^n\times\R^n,\quad \widetilde{\bfx}\mapsto \left(\bfPhi(\widetilde{\bfx}), \underline{\bfPhi}(\widetilde{\bfx}), \overline{\bfPhi}(\widetilde{\bfx})\right)
\end{equation}
where $\underline{\bfPhi}$ and $\overline{\bfPhi}$ are mappings to lower and upper interval bounds for the prediction of the \IntervalShort, cf.\ supplementary material. Given labeled samples $(\widetilde{\bfx}_i, \bfx_i)=(\bfA^\dagger\bfy_i, \bfx_i)$ it is suggested in \cite{INN} to train the \IntervalShort\ by minimizing the empirical loss
\begin{equation*}\label{eq:interval-loss}
 \textstyle\sum_i\|\max \{\bfx_i-\overline{\bfPhi}(\widetilde{\bfx}_i),0\}\|_2^2 + \|\max\{\underline{\bfPhi}(\widetilde{\bfx}_i)-\bfx_i,0\}\|_2^2 + \beta\|\overline{\bfPhi}(\widetilde{\bfx}_i)-\underline{\bfPhi}(\widetilde{\bfx}_i)\|_1,
\end{equation*}
subject to constraints that guarantee $\underline{\bfPhi}(\widetilde{\bfx})\leq\bfPhi(\widetilde{\bfx})\leq\overline{\bfPhi}(\widetilde{\bfx})$ for all $\widetilde{\bfx}$. Hence, the idea of \IntervalShort s is to produce output intervals that contain the true labels with high probability, while remaining as tight as possible. The pixel-wise uncertainty estimate of an \IntervalShort\ is then given by the width of the prediction interval, i.e., $\bfu_{\text{\IntervalShort}}(\widetilde{\bfx}) = \overline{\bfPhi}(\widetilde{\bfx}) - \underline{\bfPhi}(\widetilde{\bfx})$. We refer to~\cite{INN} for further details on \IntervalShort s and their evaluation in the context of uncertainty quantification.

\subsubsection{\DropLong} In \DropShort\ proposed by \cite{gal_dropout_2016,kendall_what_2017}, uncertainty scores are obtained through the sample variance of multiple stochastic forward passes on the same input data point. In other words, if $\bfPhi_1,\dots,\bfPhi_T$ are realizations of independent draws of random dropout masks for the same prediction network $\bfPhi$, then the pixel-wise uncertainty estimate is given by
\begin{equation*}
\textstyle\bfu_{\text{\DropShort}}(\widetilde{\bfx}) = \frac{1}{T-1}\left( \sum_{t=1}^T \bfPhi_t(\widetilde{\bfx})^2- \frac{1}{T} \left(\sum_{t=1}^T \bfPhi_t(\widetilde{\bfx})\right)^2\right).
\end{equation*}
\subsubsection{\ProbLong}
The work by \cite{nix_estimating_1994} proposed another simple recipe for uncertainty scores: the number of output components of the prediction network is doubled and trained to approximate the mean and variance of a Gaussian distribution. This approach has been recast by \cite{gast_lightweight_2018} as so-called lightweight probabilistic networks (\ProbShort)
\[
\textstyle\bfPhi_{\text{\ProbShort}}\colon\R^n\to\R^n\times\R^n,\quad \widetilde{\bfx}\mapsto (\bfPhi_{\text{mean}}(\widetilde{\bfx}), \bfPhi_{\text{var}}(\widetilde{\bfx}))
\]
which are trained by minimizing the empirical loss
\begin{equation*}
    \textstyle\sum_i \left\|\frac{\bfx_i-\bfPhi_{\text{mean}}(\widetilde{\bfx}_i)}{\sqrt{\bfPhi_{\text{var}}(\widetilde{\bfx}_i)}}\right\|_2^2 + \|\log \bfPhi_{\text{var}}(\widetilde{\bfx}_i)\|_1.
\end{equation*}
The pixel-wise uncertainty score of \ProbShort\ is then simply given by the variance estimate, i.e., $\bfu_{\text{\ProbShort}}(\widetilde{\bfx}) = \bfPhi_{\text{var}}(\widetilde{\bfx})$.
\section{Experiments and Results}
\label{sec:main}
In this section, we first briefly report on the general deep learning setup of the experiments. Detailed technicalities are listed in the supplement for the sake of reproducability. Finally, we describe the actual experiments for the detection of instabilities and their results. 

\subsection{Inverse Problems, Neural Networks and Data}
\label{sec:architectures}
\subsubsection{\Denoise}
This task consists of removing additive Gaussian noise with standard deviation $25/255$ from greyscale images (rescaled to the intensity range $[0,1]$) from the Berkeley Segmentation Dataset \cite{martin2001bsd}. The prediction network underlying all uncertainty methods is a fully-convolutional residual network with 17 convolution layers, inspired by \cite{Zhang2017}.

\input{figures/bsd68/bsd_adv_ood.tex}

\subsubsection{Limited Angle Computed Tomography (CT)}
For this task, we consider a simulation of the noiseless Radon transform with a moderate missing wedge of $30^\circ$ for the forward model~\eqref{eq:inv_prob}. The non-learned inversion $\bfA^\dagger$ in~\eqref{eq:inv_method} is based on the filtered backprojection algorithm (FBP)~\cite{natterer2001}. The underlying prediction network is a U-Net~\cite{ronneberger_u-net:_2015} variant. Our experiments are based on a data set consisting of $512\times 512$ human CT scans from the AAPM Low Dose CT Grand Challenge data \cite{mayo}.\footnote{See:~\url{https://www.aapm.org/GrandChallenge/LowDoseCT/}; We would like to thank Dr.\ Cynthia McCollough, the
Mayo Clinic, and the American Association of Physicists in Medicine as well as the grants EB017095 and
EB017185 from the National Institute of Biomedical Imaging and Bioengineering for providing the AAPM data.} In total, it contains $2580$ images of $10$ patients. Eight of these ten patients were used for training ($2036$ samples), one for validation ($214$ samples) and one for testing ($330$ samples).

\subsection{Instability Detection}
\label{sec:experiments}
Two experiments are performed on the two tasks described above. The first one, \AdvLong, examines the capacity of uncertainty quantification methods to detect adversarial inputs. The second experiment, \ArtLong, exposes the prediction network to a novel structure that was not present during training, analogous to the out-of-distribution test in \cite{vegard2019,gottschling2020}. Both experiments are explained in detail below.
\subsubsection{\AdvLong\ (\AdvShort)} 
The \AdvShort\ experiment assesses the capacity of the considered UQ methods to capture artifacts in the output that were caused by adversarial noise. To that end, we create perturbed inputs for each measurement sample $\bfy$ in the test set by employing the box-constrained L-BFGS algorithm \cite{10.1137/0916069} to solve
\begin{equation}\label{eq:adv_opt}
    \textstyle
    \minimize_{\widetilde{\bfx}_{\text{adv}}\in[0,1]^n} \|\bfPhi(\widetilde{\bfx}_{\text{adv}})-\bfx_{\text{adv.~tar.}}\|_2^2 + \lambda\|\widetilde{\bfx}_{\text{adv}}-\widetilde{\bfx}\|_2^2,
\end{equation}
where $\widetilde{\bfx} = \bfA^\dagger \bfy$ denotes the model based inversion, $\bfx_{\text{adv.~tar.}}$ represents a corresponding adversarial target, and $\lambda\geq 0$ is a parameter for balancing the two terms in~\eqref{eq:adv_opt}. It is arguable, whether the technical aspects of such an adversarial pertubation (i.e., attacking subsequently to a model-based inversion) is a realistic scenario in the context of inverse problems. However, for our purposes, such a simple setup (see also~\cite{huang2018}) is sufficient. We refer to~\cite{vegard2019,gottschling2020}, where adversarial noise is mapped to the measurement domain. 
For the \Denoise\ data we use $\lambda=0.5$, and the adversarial targets are created by adding noise to a random $50\times 50$ patch in the reconstruction $\bfx_{\text{rec}} = \bfPhi(\widetilde{\bfx})$ obtained via~\eqref{eq:inv_method}. Thus, the denoising network is forced to fail its task in that region; see \Cref{fig:bsd_adv}. For the \ChestCT\ task we found that the second term in \eqref{eq:adv_opt} is not required, i.e., we use $\lambda=0$. Adversarial targets are created by subtracting $1.5$ times its mean value from $\bfx_{\text{rec}}$ within a random $50\times 50$ square, leading to clearly visible artifacts in the corresponding reconstructions; see \Cref{fig:adv_1}. In order to assess the adversarial artifact detection capacity, the different UQ schemes are then used to produce uncertainty heatmaps for the generated adversarial inputs. A quantitative evaluation is carried out by computing the mean Pearson correlation coefficient between the pixel-wise change in the uncertainty heatmaps $|\bfu(\widetilde{\bfx})-\bfu(\widetilde{\bfx}_{\text{adv}})|$ and the change of reconstructions $|\bfx_{\text{rec}}-\bfPhi(\widetilde{\bfx}_{\text{adv}})|$. The results are summarized in \Cref{tab:results} and illustrated in \Cref{fig:bsd_adv,fig:adv_1}. We observe that both \IntervalShort\ and \ProbShort\ are able to detect the image region of adversarial perturbations, with \ProbShort\ achieving slightly higher correlations in the denoising task and \IntervalShort\ having the highest correlation in the \CTShort\ task. This shows that both methods are able to visually highlight the effect that almost imperceptible input perturbations have on the reconstructions.  

\subsubsection{\ArtLong\ (\ArtShort)}
The \ArtShort\ experiment is designed analogous to the setup described by \cite{gottschling2020}, i.e., an atypical artifact, which was not present in the training data, is randomly placed in the input. For the \Denoise\ task this is achieved by locally changing the noise distribution, i.e., we replace the Gaussian noise by Salt \& Pepper noise in one half of each image in the test set; see \Cref{fig:bsd_adv}. For the \ChestCT\ task the silhouette of a peace dove is inserted in each image of the test set; see \Cref{fig:adv_1}. The simulation of the measurements and model-based inversions is carried out on the new test set as before.
In order to assess the atypical artifact detection capacity, the different UQ schemes are then used to produce uncertainty heatmaps on the resulting OoD inputs. A quantitative evaluation is carried out by computing the mean Pearson correlation coefficient between the change in the uncertainty heatmaps $|\bfu(\widetilde{\bfx})-\bfu(\widetilde{\bfx}_{\text{OoD}})|$ and a binary mask marking the region of change in the inputs. The results are summarized in \Cref{tab:results} and illustrated in \Cref{fig:bsd_adv,fig:adv_1}. All three UQ methods are correlated with the input change, however \IntervalShort\ achieves the highest correlation in both the \Denoise\ and \CTShort\ task. This shows that UQ in general, and \IntervalShort s in particular, can serve as a warning system for inputs containing atypical features that might otherwise lead to unnoticed and possibly erroneous reconstruction artifacts.
\section{Conclusion}
\label{sec:conclusion}
We demonstrated qualitatively and quantitatively on two use-cases, image denoising and limited angle computed tomography, that uncertainty quantification, in particular \IntervalShort \ and \ProbShort, bears great potential as a fine-grained instability detector. Furthermore, \IntervalLong s performed best overall in three out of four experiments. The implication and goal of this work is to ultimately move deep learning technology closer to a level of reliability that makes it a serious contender for integration in medical imaging workflows. If we want to harness the prowess of deep learning we will need to find strategies for accounting for its instabilities. Uncertainty quantification can be an important tool to that end.

\begin{table}
\caption{
Mean Pearson correlation coefficients, averaged ($\pm$ standard deviation) over three experimental runs, for the \AdvLong \ and \ArtLong \ experiments.}
\label{tab:results}
\begin{center}
\begin{small}

\begin{tabular}{l@{\qquad}c@{\;\;}c@{\qquad}c@{\;\;}c}
\toprule
 & \multicolumn{2}{c}{\Denoise} 
 & \multicolumn{2}{c}{\CTShort} \\
 
   UQ Method 
 & \AdvShort 
 & \ArtShort 
 & \AdvShort  
 & \ArtShort \\
 
 \midrule

   \IntervalShort 
 & $0.77\pm0.008$ 
 & $\mathbf{0.69\pm0.006}$ 
 & $\mathbf{0.56\pm0.05}$ 
 & $\mathbf{0.52\pm0.03}$ \\
 
   \DropShort
 & $0.20\pm0.001$
 & $0.44\pm0.02$
 & $0.28\pm0.02$
 & $0.26\pm0.01$ \\
 
  \ProbShort
& $\mathbf{0.81\pm0.002}$
& $0.44\pm0.01$
& $0.48\pm0.12$
& $0.34\pm0.04$\\

 \bottomrule
\end{tabular}

\end{small}
\end{center}
\end{table}

\input{figures/ct240_contrast/ct240_contrast}
\newpage

\appendix

\section{Supplementary Material}

\subsection{Details of Experimental Setup}

\begin{table}
\centering
\caption{Summary of the technical details regarding the neural network architecures, training, and data sets for the two use cases of \Denoise\ and \CTLong. \Denoise\ data is available at \url{https://github.com/husqin/DnCNN-keras}(not affiliated with authors of this paper).}
\begin{tabular}{lll}
\toprule
 & \textbf{Image Denoising} & \textbf{Limited Angle CT} \\
 \midrule
 \multirow{7}{*}{\rotatebox[origin=c]{90}{\textbf{Base Network}}} & based on \cite{Zhang2017}  & U-Net of~\cite{ronneberger_u-net:_2015} \\
 & dropout (0.05) after every other conv. & dropout (0.7) after down-/up-sampling \\
  & trained with Adam\cite{kingma2014adam}, 50 epochs & trained with Adam, 400 epochs \\
  & learning rate: $10^{-4}$ & learning rate: $7.5\cdot 10^{-5}$ \\
  & mini-batch size: 128 & mini-batch size: 12\\
   & no batch normalization as in \cite{Zhang2017} &  \\
 & 128 instead of 64 conv.~channels, cf.\ \cite{Zhang2017} & \\
  \midrule
\multirow{5}{*}{\rotatebox[origin=c]{90}{\textbf{\IntervalShort}}} & 10 epochs with Adam  &  15 epochs with Adam \\
& learning rate: $10^{-6}$ & learning rate: $10^{-6}$\\
& $\beta = 10^{-3}$ & $\beta=10^{-4}$ \\
& mini batch size: 96 & mini batch size: 6\\
& interval arithmetic in last 8 layers & interval arithmetic in last 12 layers \\
\midrule
\rotatebox[origin=c]{90}{\textbf{\DropShort}}  & $T=128$ forward passes   & $T=16$ forward passes\\
\hline
 \multirow{4}*{\rotatebox[origin=c]{90}{\textbf{\ProbShort}}} & additional output channel &  additional output channel \\
 & otherwise same setup as base network & 400 more epochs with Adam \\
 & & learning rate: $10^{-7}$\\
 & & mini-batch size: 12 \\
 \midrule
\multirow{6}*{\rotatebox[origin=c]{90}{\textbf{Data}}} & Berkeley Segmentation Dataset~\cite{martin2001bsd} & AAPM Low Dose CT Grand Challenge \\
& 400 $128\times 128$-images; see~\cite{schmidt2014shrinkagefields,Zhang2017} & 10 patients: 2580 $512\times 512$-images \\
& overlapping $40\times 40$-patches, stride 10 & (8/1/1 for training/validation/testing) \\
& rescaled to intesity range $[0,1]$ & noiseless Radon transform \\
& Gaussian noise, standard dev.~$25/255$ &  $30^\circ$ missing wedge\\
& testing: 68 images of varying size; cf.\ \cite{Zhang2017} & Ramp-filter for FBP\\
\bottomrule
\vspace{-1cm}
\end{tabular}

\end{table}

\subsection{Interval Arithmetic in Neural Networks}
\begin{figure*}[ht]
    \centering
    \input{graphs/tikz_schemaic}
    \vspace{-0.7cm}
    \caption{\textbf{\IntervalShort\ Schematic Overview.} The structure of an \IntervalLong\, figure reproduced from \cite{INN} with permission from the authors.}
    \label{fig:interval_overview}
    \vspace{-0.5cm}
\end{figure*}

We give a derivation of the lower and upper interval bounds $\underline{\bf\Phi}$ and $\overline{\bfPhi}$ in equation (3) of the main paper.
\IntervalLong s (\IntervalShort s) make use of interval arithmetic that deviates from customary arithmetic. The forward pass through a ReLU neural network layer $\bfx\mapsto \varrho(\bfW\bfx+\bfb)$ in interval arithmetic is as follows: Given a component-wise interval valued input $[\underline{\bfx}, \overline{\bfx}]$ and interval valued weight matrices $\left[\underline{\bfW}, \overline{\bfW}\right]$ and bias vectors $\left[\underline{\bfb}, \overline{\bfb}\right]$ the output interval $[\underline{\bfz}, \overline{\bfz}]$ after propagation through the layer is formally expressed as
\begin{equation*}
\begin{aligned}
    \left[\underline{\bfz}, \overline{\bfz}\right] = \varrho
    \left(\left[\underline{\bfW}, \overline{\bfW}\right]
    \left[\underline{\bfx}, \overline{\bfx}\right]+
    \left[\underline{\bfb}, \overline{\bfb}\right] \right).
\end{aligned}\label{eq:interval_prop}
\end{equation*}
In the special case where $[\underline{\bfx}, \overline{\bfx}]$ is non-negative---for example image inputs scaled to the intensity range $[0, 1]$ or outputs of a previous ReLU layer---this can be explicitly calculated via
\begin{align*}
    \underline{\bfz} &= \varrho \left( \max \left\{ \underline{\bfW}, 0\right\}
    \underline{\bfx}+\min \left\{ \underline{\bfW}, 0\right\}
    \overline{\bfx}+\underline{\bfb} \right),\\
    \overline{\bfz} &= \varrho \left(\min \left\{ \overline{\bfW},0\right\} \underline{\bfx}+\max \left\{ \overline{\bfW}, 0 \right\} \overline{\bfx}+\overline{\bfb}\right),
\end{align*}
where the maximum and minimum functions are applied component-wise. Applying this for all network layers finally yields $\underline{\bf\Phi}$ and $\overline{\bfPhi}$.
%
%
%
\newpage
\noindent\textbf{Acknowledgements} The authors would like to thank Sören Becker for feedback on the final draft. M.M. acknowledges support by the DFG Priority Programme DFG-SPP 1798 Grants KU
1446/21 and KU 1446/23.
\bibliography{ref/new_refs}
\bibliographystyle{splncs04}

\end{document}

%% file: figures/bsd68/bsd_adv_ood.tex
\newcommand{\nozoom}[1]{%
  \raisebox{-.5\height}{\includegraphics[width=2.7cm]{#1}}%
}

\newcommand{\withzoom}[1]{%
  \raisebox{-.5\height}{\noindent
    \begin{tikzpicture}[inner sep=0, spy using outlines={rectangle,black,ultra thin,magnification=1.75,width=1.1cm,height=0.9cm}]
      \node (img) {\includegraphics[width=2.7cm]{#1}};
      \spy on (.95,0.05) in node[below right] at (img.north west);
    \end{tikzpicture}%
  }%
}

\begin{figure}
    \centering\scriptsize
    \begin{tabular}{@{}cc@{\;}cccc@{}}
        
        & & & \IntervalShort & \DropShort & \ProbShort \\
    
          \rotatebox[origin=c]{90}{input $\widetilde{\bfx}$}
        & \withzoom{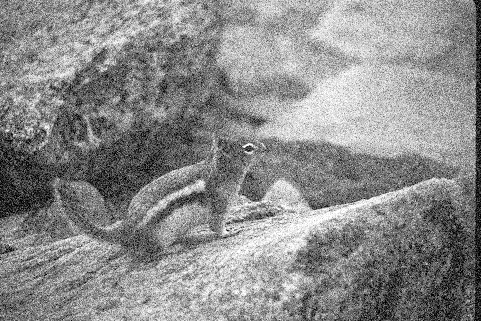}
        & \rotatebox[origin=c]{90}{\parbox{2.1cm}{\centering adv. input\\ $\widetilde{\bfx}_{\text{adv}}$}}
        & \withzoom{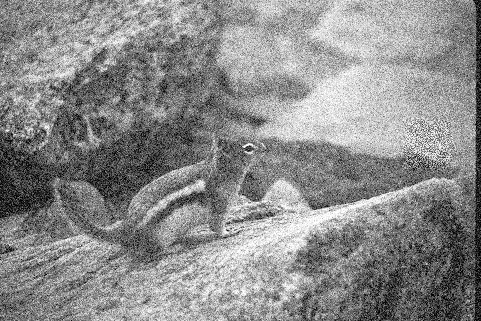}
        & \withzoom{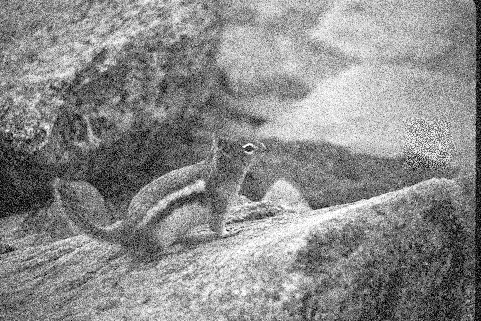}
        & \withzoom{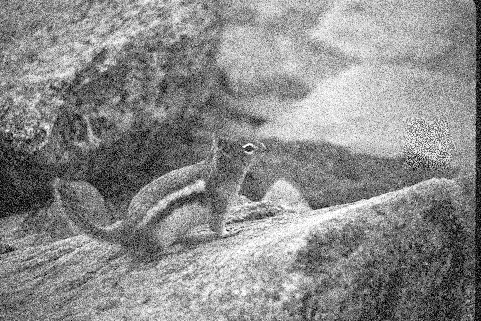} \\
        
          \rotatebox[origin=c]{90}{target $\bfx$}
        & \withzoom{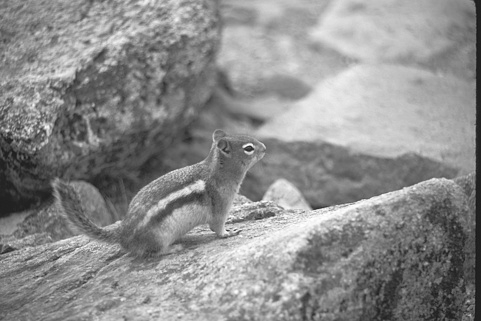}
        & \rotatebox[origin=c]{90}{\parbox{2.1cm}{\centering adv. rec.\\ $\bfPhi(\widetilde{\bfx}_{\text{adv}})$}}
        & \withzoom{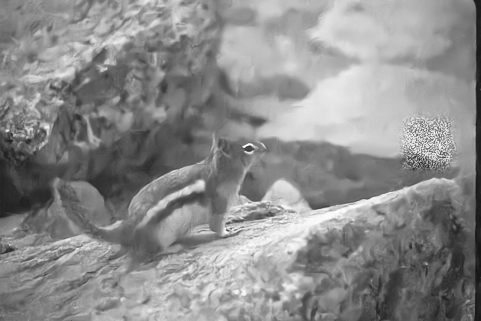}
        & \withzoom{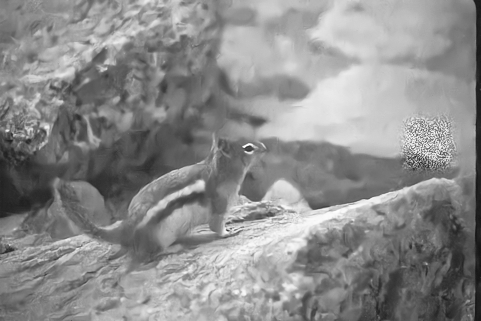}
        & \withzoom{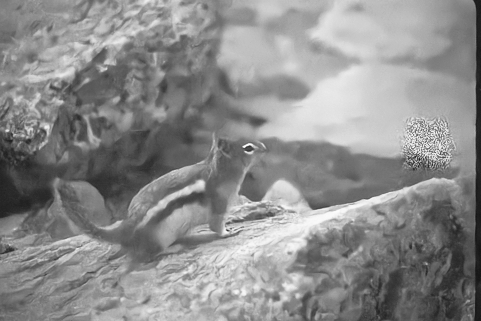} \\        
        &
        & \rotatebox[origin=c]{90}{\parbox{2.1cm}{\centering adv. uncert.\\ $\bfu(\widetilde{\bfx}_{\text{adv}})$}}
        & \withzoom{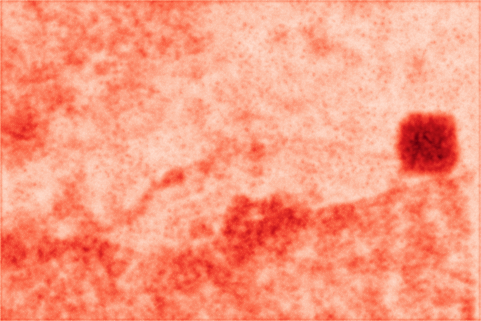}
        & \withzoom{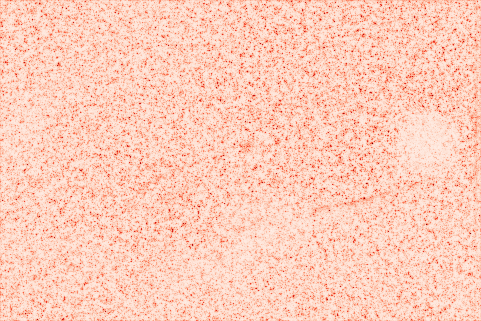}
        & \withzoom{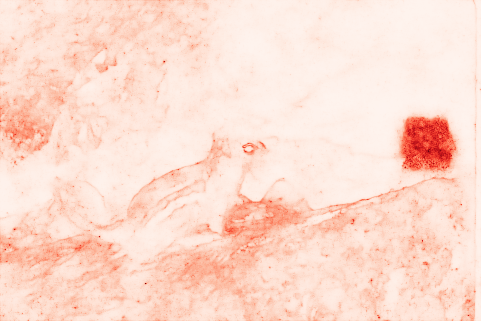} \\        
        &
        & \rotatebox[origin=c]{90}{\parbox{2.1cm}{\centering rec. difference\\ $|\bfx_{\text{rec}}-\bfPhi(\widetilde{\bfx}_{\text{adv}})|$}}
        & \withzoom{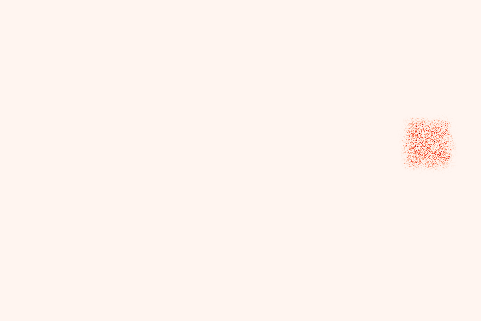}
        & \withzoom{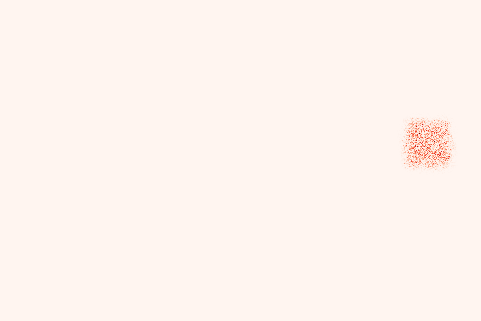}
        & \withzoom{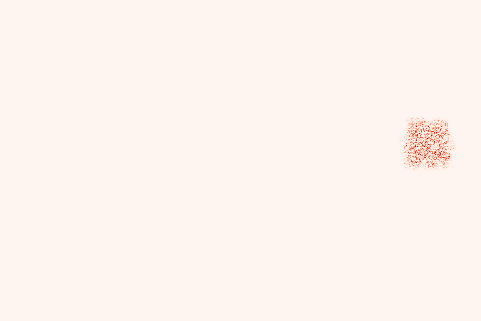} \\ \\
        
        \rotatebox[origin=c]{90}{OoD input $\widetilde{\bfx}_{\text{OoD}}$}
        & \nozoom{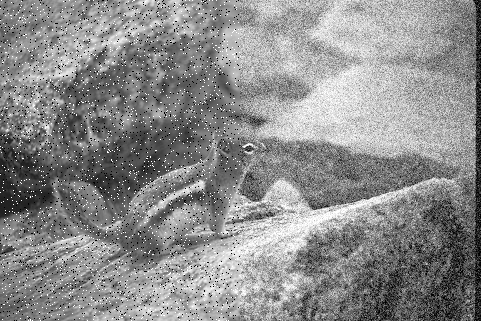}
        & \rotatebox[origin=c]{90}{\parbox{2.1cm}{\centering OoD rec.\\ $\bfPhi(\widetilde{\bfx}_{\text{OoD}})$}}
        & \nozoom{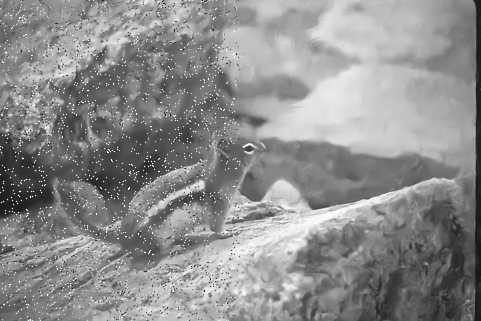}
        & \nozoom{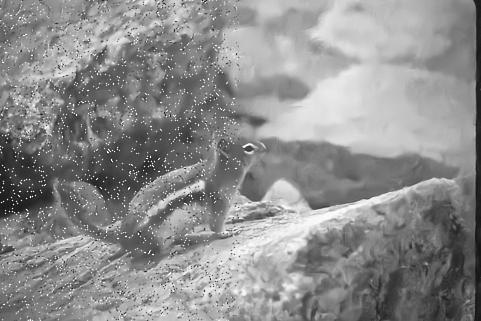}
        & \nozoom{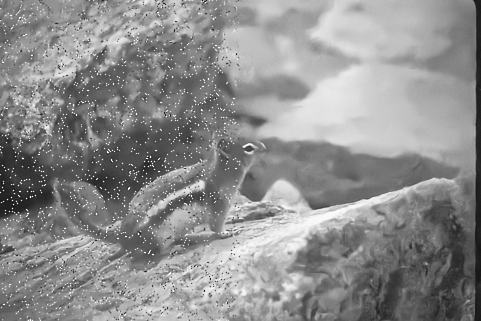} \\
        
        & 
        & \rotatebox[origin=c]{90}{\parbox{2.1cm}{\centering OoD uncert.\\ $\bfu(\widetilde{\bfx}_{\text{OoD}}$)}}
        & \nozoom{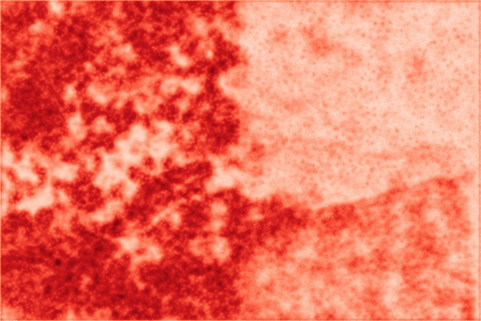}
        & \nozoom{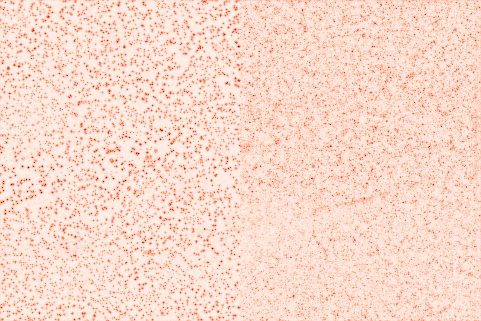}
        & \nozoom{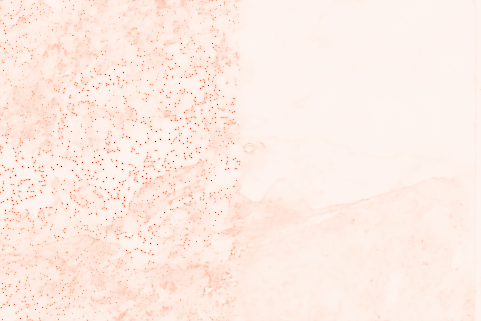}

    \end{tabular}
    
    \caption{Results of three UQ methods for the \AdvShort\ and \ArtShort\ experiments for one exemplary data sample of the \Denoise\ task.}
    \label{fig:bsd_adv}
\end{figure}

%% file: figures/ct240_contrast/ct240_contrast.tex
\newcommand{\nozoomct}[1]{%
  \raisebox{-.5\height}{\includegraphics[width=2.7cm]{#1}}%
}

\newcommand{\withzoomct}[1]{%
  \raisebox{-.5\height}{\noindent
    \begin{tikzpicture}[inner sep=0, spy using outlines={rectangle,black,ultra thin,magnification=1.75,width=1.1cm,height=0.9cm}]
      \node (img) {\includegraphics[width=2.7cm]{#1}};
      \spy on (.6,.15) in node[below right] at (img.north west);
    \end{tikzpicture}%
  }%
}

\newcommand{\withzoomctood}[1]{%
  \raisebox{-.5\height}{\noindent
    \begin{tikzpicture}[inner sep=0, spy using outlines={rectangle,black,ultra thin,magnification=1.5,width=1.2cm,height=1.0cm}]
      \node (img) {\includegraphics[width=2.7cm]{#1}};
      \spy on (-.4,-.15) in node[below left] at (img.north east);
    \end{tikzpicture}%
  }%
}

\begin{figure}
    \centering\scriptsize
    \begin{tabular}{@{}cc@{\;}cccc@{}}
        
        & & & \IntervalShort & \DropShort & \ProbShort \\
    
          \rotatebox[origin=c]{90}{input $\widetilde{\bfx}$}
        & \withzoomct{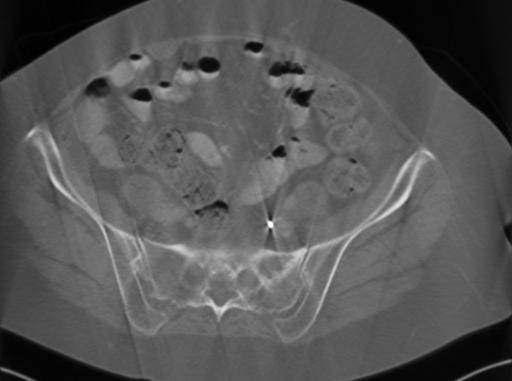}
        & \rotatebox[origin=c]{90}{\parbox{2.0cm}{\centering adv. input\\ $\widetilde{\bfx}_{\text{adv}}$}}
        & \withzoomct{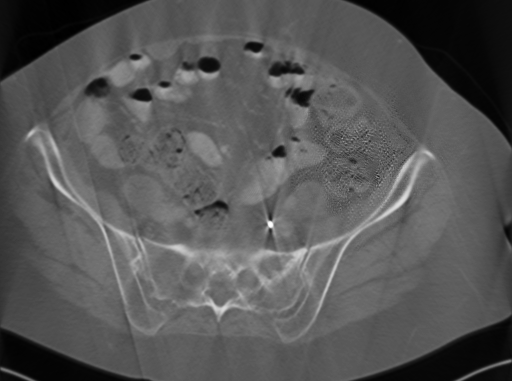}
        & \withzoomct{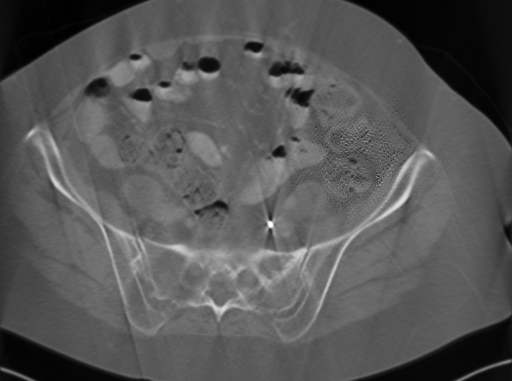}
        & \withzoomct{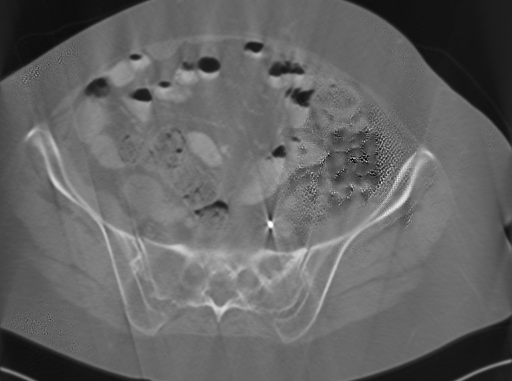} \\
        
        
        \rotatebox[origin=c]{90}{target $\bfx$}
        & \withzoomct{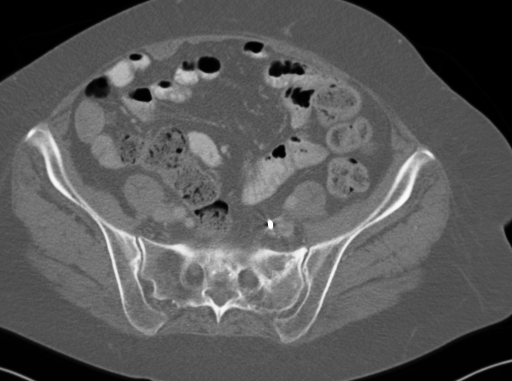}
        & \rotatebox[origin=c]{90}{\parbox{2.0cm}{\centering adv. rec.\\ $\bfPhi(\widetilde{\bfx}_{\text{adv}})$}}
        & \withzoomct{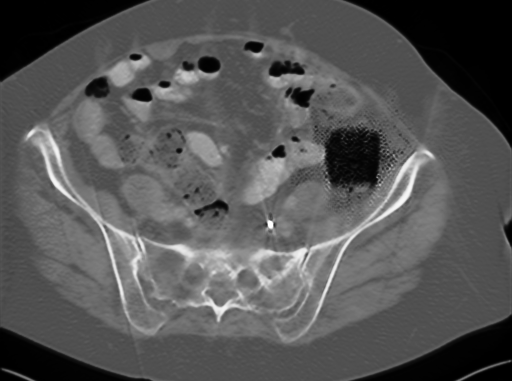}
        & \withzoomct{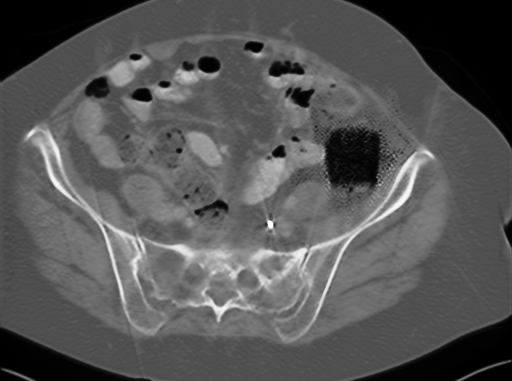}
        & \withzoomct{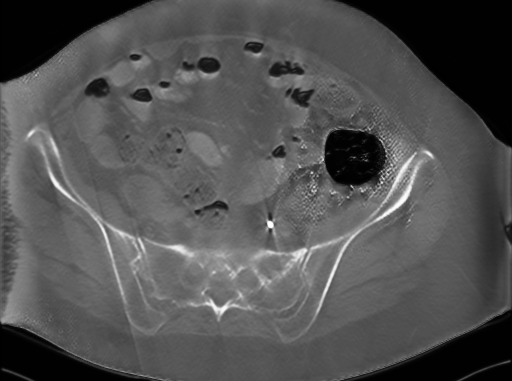} \\        
        &
        & \rotatebox[origin=c]{90}{\parbox{2.0cm}{\centering adv. uncert.\\ $\bfu(\widetilde{\bfx}_{\text{adv}})$}}
        & \withzoomct{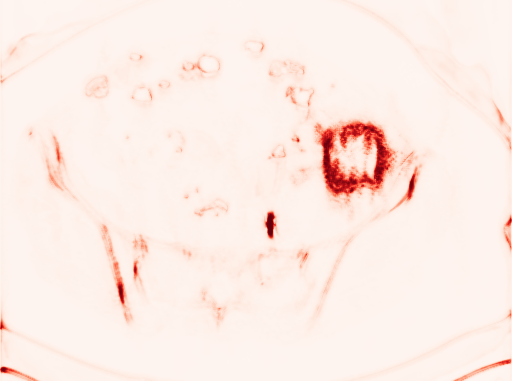}
        & \withzoomct{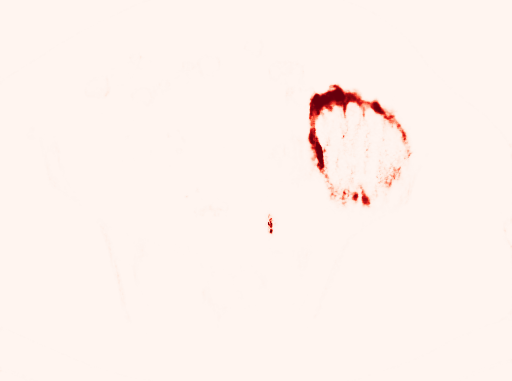}
        & \withzoomct{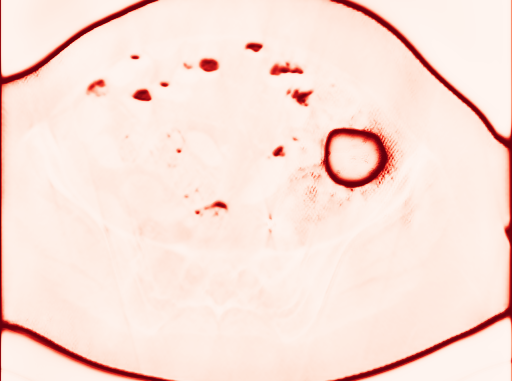} \\       
        &
        & \rotatebox[origin=c]{90}{\parbox{2.0cm}{\centering rec. difference\\ $|\bfx_{\text{rec}}-\bfPhi(\widetilde{\bfx}_{\text{adv}})|$}}
        & \withzoomct{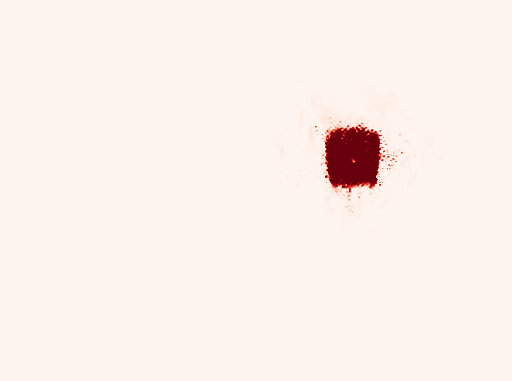}
        & \withzoomct{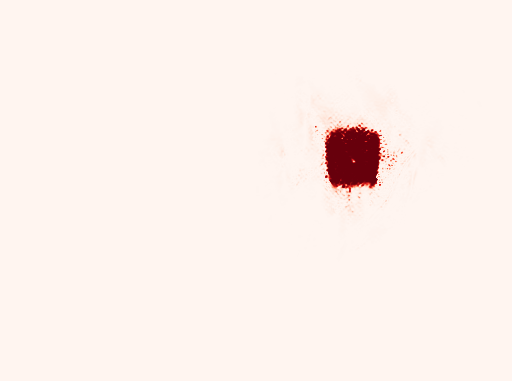}
        & \withzoomct{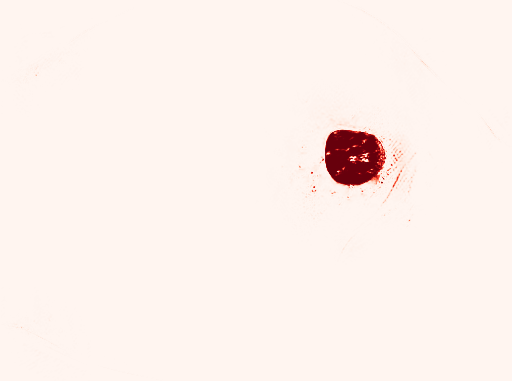} \\ \\
        
        \rotatebox[origin=c]{90}{OoD input $\widetilde{\bfx}_{\text{OoD}}$}
        & \withzoomctood{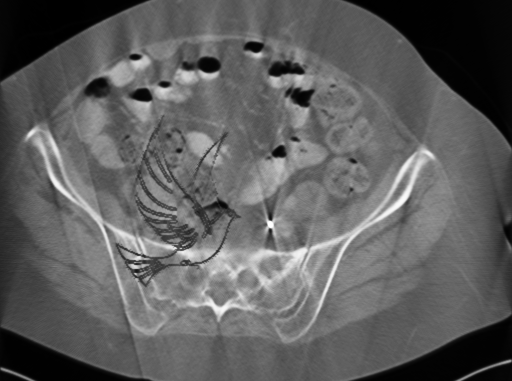}
        & \rotatebox[origin=c]{90}{\parbox{2.0cm}{\centering OoD rec.\\ $\bfPhi(\widetilde{\bfx}_{\text{OoD}})$}}
        & \withzoomctood{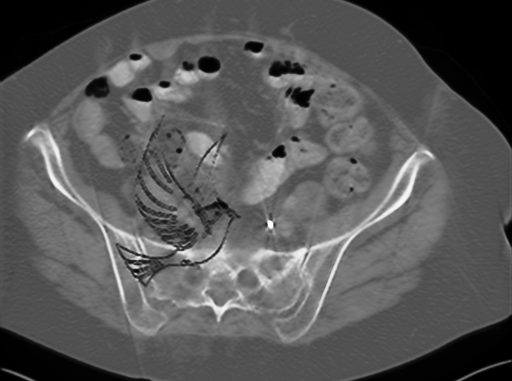}
        & \withzoomctood{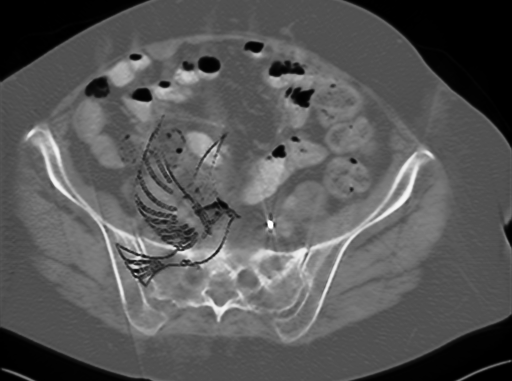}
        & \withzoomctood{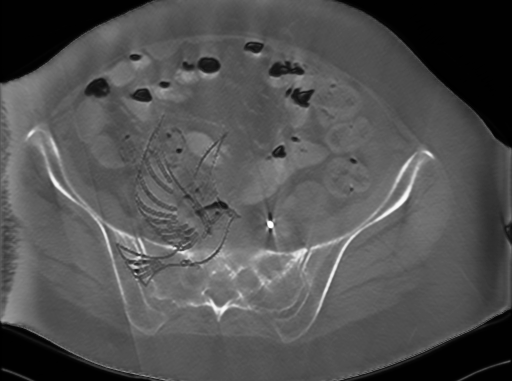} \\
        
        & 
        & \rotatebox[origin=c]{90}{\parbox{2.0cm}{\centering OoD uncert.\\ $\bfu(\widetilde{\bfx}_{\text{OoD}}$)}}
        & \withzoomctood{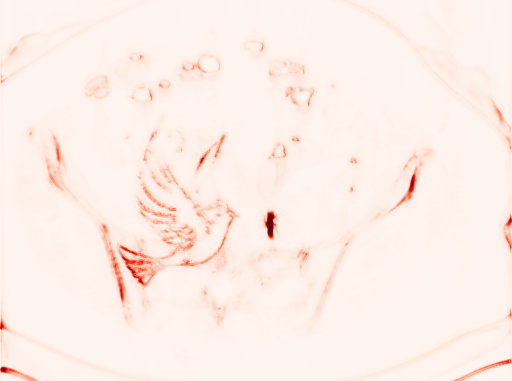}
        & \withzoomctood{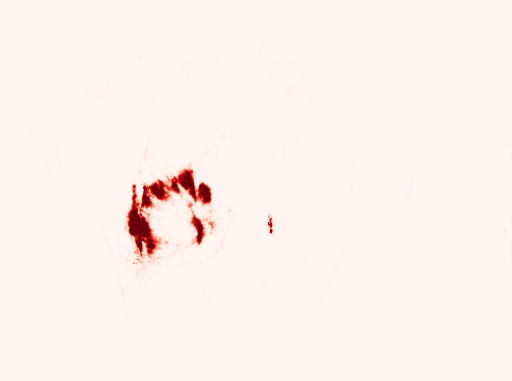}
        & \withzoomctood{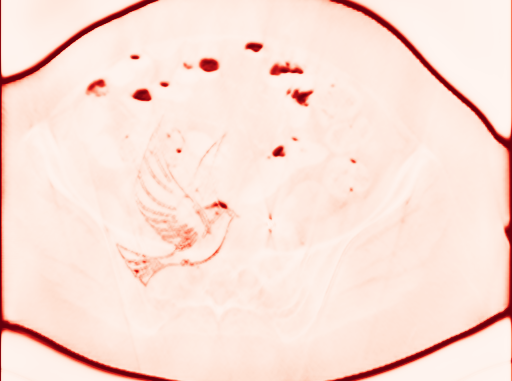}

    \end{tabular}
    \caption{Results of three UQ methods for the \AdvShort\ and \ArtShort\ experiments for one exemplary data sample of the \CTLong\ task. The plotting windows are slightly adjusted for better contrast.}
    \label{fig:adv_1}
\end{figure}

%% file: graphs/tikz_schemaic.tex
\scalebox{0.7}{
    \large
    \begin{tikzpicture}
    	\node[] at (0.3,0) {\includegraphics[scale=0.2]{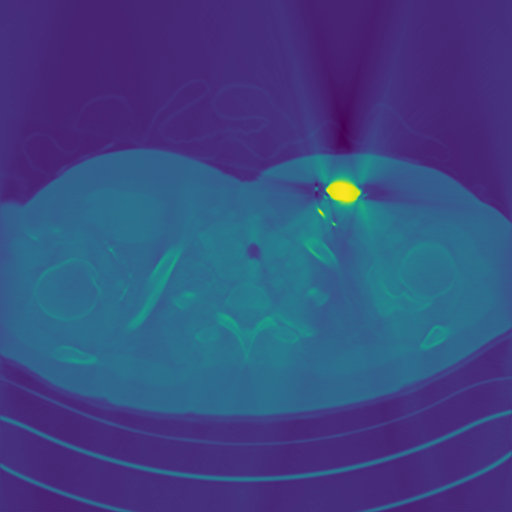}};
    	\node[] at (0.3,2) {Input};
    	
    	\node[] at (6,2) {$\text{INN}_{\color{blue}{\underline{\theta}},\color{black}{\theta},\color{red}{\overline{\theta}}}$};
    	
    	\draw[very thick,->,color=red] (2.25,0.5) -- (2.7,0.5);
    	\draw[very thick,->,color=black] (2.25,0) -- (2.7,0);
    	\draw[very thick,->,color=blue] (2.25,-0.5) -- (2.7,-0.5);
    	
    	\tikzstyle{intervalnode}=[draw,rounded corners,minimum width=2em, minimum height=1.6em,black,yscale=0.9,xscale=1.7]

    	\draw[thick,gray] (3.5cm-0.5pt,1.2) -- (3.5cm+0.5pt,1.2)
    	    node[pos=0,blue,xscale=2]{$[$}
    	    node[intervalnode ]at (3.5,1.2){$|$} 
    	    node[pos=1,red,xscale=2]{$]$};
    	    
    	\draw[thick,gray] (3.5cm-0.5pt,0.4) -- (3.5cm+0.5pt,0.4)
    	    node[pos=0,blue,xscale=2]{$[$}
    	    node[intervalnode] at (3.5,0.4){$|$} 
    	    node[pos=1,red,xscale=2]{$]$};
    
    	\draw[thick,gray] (3.5cm-0.5pt,-0.4) -- (3.5cm+0.5pt,-0.4)
    	    node[pos=0,blue,xscale=2]{$[$}
    	    node[intervalnode] at (3.5,-0.4){$|$} 
    	    node[pos=1,red,xscale=2]{$]$};
    
    	\draw[thick,gray] (3.5cm-0.5pt,-1.2) -- (3.5cm+0.5pt,-1.2)
    	    node[pos=0,blue,xscale=2]{$[$}
    	    node[intervalnode] at (3.5,-1.2){$|$} 
    	    node[pos=1,red,xscale=2]{$]$};

        \draw[thick,gray] (5.25-0.4,1.) -- (5.25+0.1,1)
    	    node[pos=0,blue,xscale=2]{$[$}
    	    node[intervalnode] at (5.25,1){$|$} 
    	    node[pos=1,red,xscale=2]{$]$};
    	    
    	\draw[thick,gray] (5.25-0.2,0.2) -- (5.25+0.3,0.2)
    	    node[pos=0,blue,xscale=2]{$[$}
    	    node[intervalnode] at (5.25,0.2){$|$} 
    	    node[pos=1,red,xscale=2]{$]$};
    
    	\draw[thick,gray] (5.25-0.4,-0.6) -- (5.25+0.45,-0.6)
    	    node[pos=0,blue,xscale=2]{$[$}
    	    node[intervalnode] at (5.25,-0.6){$|$} 
    	    node[pos=1,red,xscale=2]{$]$};

    	\draw[thick,gray] (7-0.1,1.) -- (7+0.1,1)
    	    node[pos=0,blue,xscale=2]{$[$}
    	    node[intervalnode] at (7,1){$|$} 
    	    node[pos=1,red,xscale=2]{$]$};
    	    
    	\draw[thick,gray] (7-0.2,0.2) -- (7+0.1,0.2)
    	    node[pos=0,blue,xscale=2]{$[$}
    	    node[intervalnode] at (7,0.2){$|$} 
    	    node[pos=1,red,xscale=2]{$]$};
    
    	\draw[thick,gray] (7-0.4,-0.6) -- (7+0.3,-0.6)
    	    node[pos=0,blue,xscale=2]{$[$}
    	    node[intervalnode] at (7,-0.6){$|$} 
    	    node[pos=1,red,xscale=2]{$]$};

    	\draw[thick,gray] (8.75-0.1,1.2) -- (8.75+0.4,1.2)
    	    node[pos=0,blue,xscale=2]{$[$}
    	    node[intervalnode ]at (8.75,1.2){$|$} 
    	    node[pos=1,red,xscale=2]{$]$};
    	    
    	\draw[thick,gray] (8.75-0.35,0.4) -- (8.75+0.05,0.4)
    	    node[pos=0,blue,xscale=2]{$[$}
    	    node[intervalnode] at (8.75,0.4){$|$} 
    	    node[pos=1,red,xscale=2]{$]$};
    
    	\draw[thick,gray] (8.75-0.3,-0.4) -- (8.75+0.4,-0.4)
    	    node[pos=0,blue,xscale=2]{$[$}
    	    node[intervalnode] at (8.75,-0.4){$|$} 
    	    node[pos=1,red,xscale=2]{$]$};
    
    	\draw[thick,gray] (8.75-0.2,-1.2) -- (8.75+0.2,-1.2)
    	    node[pos=0,blue,xscale=2]{$[$}
    	    node[intervalnode] at (8.75,-1.2){$|$} 
    	    node[pos=1,red,xscale=2]{$]$};

    	\draw[very thick,->,color=red] (9.65,0.5) -- (10.15,1.5);
    	\draw[very thick,->,color=black] (9.65,0) -- (10.15,0);
    	\draw[very thick,->,color=blue] (9.65,-0.5) -- (10.15,-1.5);
    
        \node[] at (11.3,0.85) {Output};
    	\node[] at (11.3,-0.2) {\includegraphics[scale=0.1]{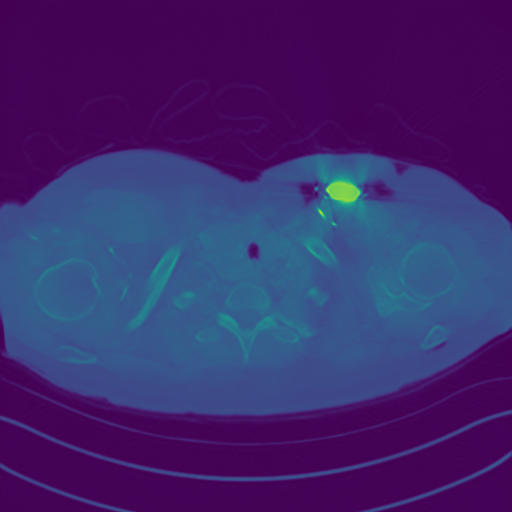}};
    	\node[] at (11.3,0.85+2.2) {Max Output};
    	\node[] at (11.3,2) {\includegraphics[scale=0.1]{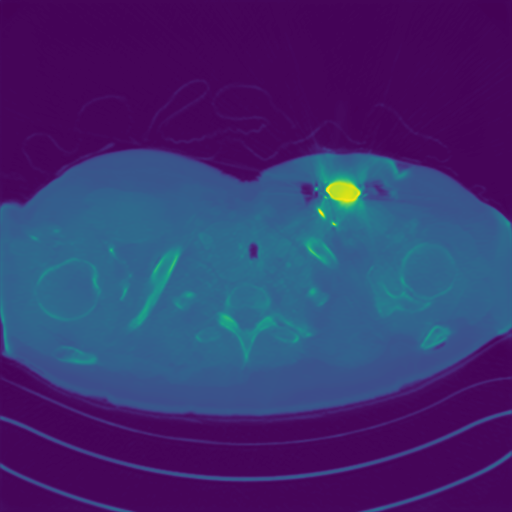}};
    	\node[] at (11.3,0.85-2.2) {Min Output};
    	\node[] at (11.3,-2.4) {\includegraphics[scale=0.1]{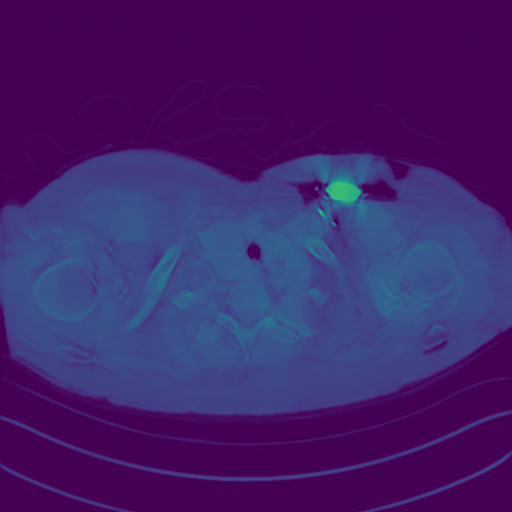}};
    	
    	\draw[very thick,->,color=red] (12.3,1.5) -- (12.6,1.2);
    	\draw[very thick,->,color=blue] (12.3,-1.5) -- (12.6,-1.2);
    	
    	\node[] at (14.05,1.5) {Uncertainty};
    	\node[] at (14.05,0) {\includegraphics[scale=0.15]{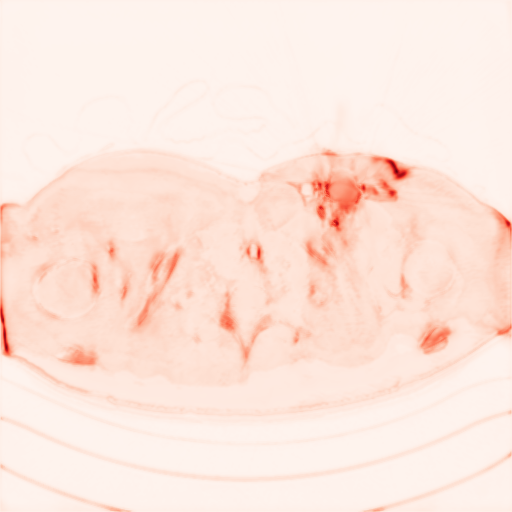}};


        \draw[very thick,black] (3.5cm+1.7em,1.2cm) -- (5.25cm-1.7em,1cm);
        
        \draw[very thick,black] (3.5cm+1.7em,1.2cm) -- (5.25cm-1.7em,-0.6cm);
        
        \draw[very thick,black] (3.5cm+1.7em,0.4cm) -- (5.25cm-1.7em,0.2cm);
        
        \draw[very thick,black] (3.5cm+1.7em,-0.4cm) -- (5.25cm-1.7em,0.2cm);
        
        \draw[very thick,black] (3.5cm+1.7em,-1.2cm) -- (5.25cm-1.7em,-0.6cm);

        \draw[very thick,black] (5.25cm+1.7em,1cm) -- (7cm-1.7em,1cm);
        
        \draw[very thick,black] (5.25cm+1.7em,1cm) -- (7cm-1.7em,0.2cm);
        
        \draw[very thick,black] (5.25cm+1.7em,-0.6cm) -- (7cm-1.7em,0.2cm);
        
        \draw[very thick,black] (5.25cm+1.7em,0.2cm) -- (7cm-1.7em,0.2cm);
        
        \draw[very thick,black] (5.25cm+1.7em,-0.6cm) -- (7cm-1.7em,1cm);
        
        \draw[very thick,black] (5.25cm+1.7em,-0.6cm) -- (7cm-1.7em,-0.6cm);

        \draw[very thick,black] (7cm+1.7em,1cm) -- (8.75cm-1.7em,1.2cm);
        
        \draw[very thick,black] (7cm+1.7em,1cm) -- (8.75cm-1.7em,0.4cm);
        
        \draw[very thick,black] (7cm+1.7em,0.2cm) -- (8.75cm-1.7em,0.4cm);
        
        \draw[very thick,black] (7cm+1.7em,-0.6cm) -- (8.75cm-1.7em,0.4cm);
        
        \draw[very thick,black] (7cm+1.7em,-0.6cm) -- (8.75cm-1.7em,-0.4cm);
        
        \draw[very thick,black] (7cm+1.7em,0.2cm) -- (8.75cm-1.7em,-1.2cm);
        
        \draw[very thick,black] (7cm+1.7em,-0.6cm) -- (8.75cm-1.7em,-1.2cm);
        
    \end{tikzpicture}
}

%% file: uq-and-instabilities.bbl
\begin{thebibliography}{10}
\providecommand{\url}[1]{\texttt{#1}}
\providecommand{\urlprefix}{URL }
\providecommand{\doi}[1]{https://doi.org/#1}

\bibitem{adler2018}
Adler, J., {\"O}ktem, O.: {Learned Primal-dual Reconstruction}. IEEE T. Med.
  Imaging  \textbf{37}(6),  1322--1332 (2018)

\bibitem{adler_deep_2018}
Adler, J., Öktem, O.: Deep {Bayesian} {Inversion}  (Nov 2018), arXiv:
  1811.05910

\bibitem{vegard2019}
Antun, V., Renna, F., Poon, C., Adcock, B., Hansen, A.C.: On instabilities of
  deep learning in image reconstruction - does {AI} come at a cost?  (2019),
  arXiv:1902.05300

\bibitem{ardizzone2018}
Ardizzone, L., Kruse, J., Wirkert, S.J., Rahner, D., Pellegrini, E.W., Klessen,
  R.S., Maier{-}Hein, L., Rother, C., K{\"{o}}the, U.: Analyzing inverse
  problems with invertible neural networks. In: International Conference on
  Learning Representations (2018)

\bibitem{arridge_maass_oektem_schoenlieb_2019}
Arridge, S., Maass, P., {\"O}ktem, O., Sch{\"o}nlieb, C.B.: Solving inverse
  problems using data-driven models. Acta Numerica  \textbf{28},  1–174
  (2019)

\bibitem{Bubba_2019}
Bubba, T.A., Kutyniok, G., Lassas, M., März, M., Samek, W., Siltanen, S.,
  Srinivasan, V.: Learning the invisible: a hybrid deep learning-shearlet
  framework for limited angle computed tomography. Inverse Problems
  \textbf{35}(6),  064002 (2019)

\bibitem{10.1137/0916069}
Byrd, R.H., Lu, P., Nocedal, J., Zhu, C.: A limited memory algorithm for bound
  constrained optimization. SIAM J. Sci. Comput.  \textbf{16}(5),  1190–1208
  (1995)

\bibitem{10.1145/3128572.3140444}
Carlini, N., Wagner, D.: Adversarial examples are not easily detected:
  Bypassing ten detection methods. In: Proceedings of the 10th ACM Workshop on
  Artificial Intelligence and Security. p. 3–14. AISec ’17 (2017)

\bibitem{Foucart2013}
Foucart, S., Rauhut, H.: A Mathematical Introduction to Compressive Sensing.
  Applied and Numerical Harmonic Analysis, Birkh{\"a}user (2013)

\bibitem{gal_dropout_2016}
Gal, Y., Ghahramani, Z.: Dropout as a bayesian approximation: Representing
  model uncertainty in deep learning. In: Balcan, M.F., Weinberger, K.Q. (eds.)
  Proceedings of The 33rd International Conference on Machine Learning.
  Proceedings of Machine Learning Research, vol.~48, pp. 1050--1059. New York,
  New York, USA (2016)

\bibitem{gast_lightweight_2018}
Gast, J., Roth, S.: Lightweight {Probabilistic} {Deep} {Networks}. 2018
  IEEE/CVF Conference on Computer Vision and Pattern Recognition pp. 3369--3378
  (2018)

\bibitem{gottschling2020}
Gottschling, N.M., Antun, V., Adcock, B., Hansen, A.C.: {The troublesome
  kernel: why deep learning for inverse problems is typically unstable?}
  (2020), arXiv:2001.01258

\bibitem{doi:10.1021/acscentsci.7b00572}
Gómez-Bombarelli, R., Wei, J.N., Duvenaud, D., Hernández-Lobato, J.M.,
  Sánchez-Lengeling, B., Sheberla, D., Aguilera-Iparraguirre, J., Hirzel,
  T.D., Adams, R.P., Aspuru-Guzik, A.: Automatic chemical design using a
  data-driven continuous representation of molecules. ACS Central Science
  \textbf{4}(2),  268--276 (2018)

\bibitem{Hammernik2018}
Hammernik, K., Klatzer, T., Kobler, E., Recht, M.P., Sodickson, D.K., Pock, T.,
  Knoll, F.: Learning a variational network for reconstruction of accelerated
  mri data. Magnetic Resonance in Medicine  \textbf{79}(6),  3055--3071 (2018)

\bibitem{hansen2010invprobs}
Hansen, P.C.: Discrete Inverse Problems. Society for Industrial and Applied
  Mathematics (2010)

\bibitem{hendrycks2019oe}
Hendrycks, D., Mazeika, M., Dietterich, T.: Deep anomaly detection with outlier
  exposure. Proc. of the International Conference on Learning Representations
  (2019)

\bibitem{huang2018}
Huang, Y., W{\"u}rfl, T., Breininger, K., Liu, L., Lauritsch, G., Maier, A.:
  Some investigations on robustness of deep learning in limited angle
  tomography. In: Frangi, A.F., Schnabel, J.A., Davatzikos, C.,
  Alberola-L{\'o}pez, C., Fichtinger, G. (eds.) Medical Image Computing and
  Computer Assisted Intervention -- MICCAI 2018. pp. 145--153 (2018)

\bibitem{jin_deep_2017}
Jin, K.H., McCann, M.T., Froustey, E., Unser, M.: Deep {Convolutional} {Neural}
  {Network} for {Inverse} {Problems} in {Imaging}. IEEE Trans. Imag. Proc.
  \textbf{26},  4509--4522 (2017)

\bibitem{kang2017}
Kang, E., Min, J., Ye, J.C.: {A deep convolutional neural network using
  directional wavelets for low-dose X-ray CT reconstruction}. Med. Phys.
  \textbf{44}(10),  360--375 (2017)

\bibitem{kendall_what_2017}
Kendall, A., Gal, Y.: What uncertainties do we need in bayesian deep learning
  for computer vision? In: Proceedings of the 31st International Conference on
  Neural Information Processing Systems. p. 5580–5590. NIPS’17, Curran
  Associates Inc., Red Hook, NY, USA (2017)

\bibitem{kingma2014adam}
Kingma, D.P., Ba, J.A.: A method for stochastic optimization. arXiv preprint
  arXiv:1412.6980  (2014)

\bibitem{LiangLS17}
Liang, S., Li, Y., Srikant, R.: Principled detection of out-of-distribution
  examples in neural networks. Proceedings of the International Conference on
  Learning Representations  (2019)

\bibitem{martin2001bsd}
Martin, D., Fowlkes, C., Tal, D., Malik, J.: A database of human segmented
  natural images and its application to evaluating segmentation algorithms and
  measuring ecological statistics. In: Proc. 8th Int'l Conf. Computer Vision.
  vol.~2, pp. 416--423 (July 2001)

\bibitem{mayo}
McCollough, C.: Tu-fg-207a-04: Overview of the low dose ct grand challenge.
  Med. Physs  \textbf{43}(6 Part 35),  3759--3760 (2016)

\bibitem{moosavi}
Moosavi-Dezfooli, S.M., Fawzi, A., Frossard, P.: Deepfool: A simple and
  accurate method to fool deep neural networks. pp. 2574--2582 (06 2016)

\bibitem{natterer2001}
Natterer, F.: The Mathematics of Computerized Tomography. SIAM (2001)

\bibitem{nix_estimating_1994}
Nix, D.A., Weigend, A.S.: Estimating the mean and variance of the target
  probability distribution. In: Proceedings of 1994 {IEEE} {International}
  {Conference} on {Neural} {Networks} ({ICNN}'94). vol.~1 (Jun 1994)

\bibitem{INN}
Oala, L., Heiß, C., Macdonald, J., März, M., Samek, W., Kutyniok, G.:
  Interval neural networks: Uncertainty scores (2020), arXiv:2003.11566

\bibitem{oymak}
Oymak, S., Hassibi, B.: Sharp mse bounds for proximal denoising. Found. Comput.
  Math.  \textbf{16} (2016)

\bibitem{Raj2020}
Raj, A., Bresler, Y., Li, B.: {Improving Robustness of Deep-Learning-Based
  Image Reconstruction} (2020), arXiv:2002.11821

\bibitem{ronneberger_u-net:_2015}
Ronneberger, O., Fischer, P., Brox, T.: U-{Net}: {Convolutional} {Networks} for
  {Biomedical} {Image} {Segmentation}. In: Navab, N., Hornegger, J., Wells,
  W.M., Frangi, A.F. (eds.) Medical {Image} {Computing} and
  {Computer}-{Assisted} {Intervention} – {MICCAI} 2015. pp. 234--241 (2015)

\bibitem{doi:10.1080/01621459.1984.10477105}
Rousseeuw, P.J.: Least median of squares regression. Journal of the American
  Statistical Association  \textbf{79}(388),  871--880 (1984)

\bibitem{schmidt2014shrinkagefields}
{Schmidt}, U., {Roth}, S.: Shrinkage fields for effective image restoration.
  In: 2014 IEEE Conference on Computer Vision and Pattern Recognition. pp.
  2774--2781 (2014)

\bibitem{42503}
Szegedy, C., Zaremba, W., Sutskever, I., Bruna, J., Erhan, D., Goodfellow, I.,
  Fergus, R.: Intriguing properties of neural networks. In: International
  Conference on Learning Representations (2014)

\bibitem{10.1023/B:MACH.0000008084.60811.49}
Tax, D.M.J., Duin, R.P.W.: Support vector data description. Mach. Learn.
  \textbf{54}(1),  45–66 (2004)

\bibitem{venkat2013}
{Venkatakrishnan}, S.V., {Bouman}, C.A., {Wohlberg}, B.: Plug-and-play priors
  for model based reconstruction. In: 2013 IEEE Global Conference on Signal and
  Information Processing. pp. 945--948 (2013)

\bibitem{zhang2016}
Zhang, H., Li, L., Qiao, K., Wang, L., et~al.: {Image Prediction for
  Limited-angle Tomography via Deep Learning with Convolutional Neural
  Network}. arXiv:1607.08707  (2016)

\bibitem{Zhang2017}
Zhang, K., Zuo, W., Chen, Y., Meng, D., Zhang, L.: Beyond a gaussian denoiser:
  Residual learning of deep cnn for image denoising. IEEE Trans. Imag. Proc.
  \textbf{26},  3142--3155 (2017)

\end{thebibliography}
